\documentclass[final]{aipproc}

\layoutstyle{6x9}

\begin{document}

\title{Evidence for Terrestrial Planetary System Remnants at White Dwarfs}

\classification{97.82}

\keywords{circumstellar matter---
	minor planets, asteroids---
	planetary systems --
	stars: abundances---
	stars: evolution---
	white dwarfs}
	
\author{J. Farihi}
{address={Department of Physics \& Astronomy, University of Leicester, Leicester LE1 7RH, UK; jf123@star.le.ac.uk}}

\begin{abstract}

The last several years have brought about a dynamic shift in the view of exoplanetary systems 
in the post-main sequence, perhaps epitomized by the evidence for surviving rocky planetary 
bodies at white dwarfs.  Coinciding with the launch of the {\em Spitzer Space Telescope}, both 
space- and ground-based data have supported a picture whereby asteroid analogs persist at 
a significant fraction of cool white dwarfs, and are prone to tidal disruption when passing close
to the compact stellar remnant.  The ensuing debris can produce a detectable infrared excess,
and the material gradually falls onto the star, polluting the atmosphere with heavy elements 
that can be used to determine the bulk composition of the destroyed planetary body.  

Based on the observations to date, the parent bodies inferred at white dwarfs are best described 
as asteroids, and have a distinctly rocky composition similar to material found in the inner Solar 
System.  Their minimum masses are typical of large asteroids, and can approach or exceed the 
mass of Vesta and Ceres, the two largest asteroids in the Solar System.  From the number of 
stars surveyed in various studies, the fraction of white dwarfs that host terrestrial planetary 
system remnants is at least a few percent, but likely to be in the range 20$-$30\%.  Therefore, 
A- and F-type stars form terrestrial planets efficiently, with a frequency at least as high as the 
remnants detected at their white dwarf descendants.

\end{abstract}

\maketitle

%%%%%%%%%%%%%%%%%%%%%%%%%%%%%%%%%%%%%%%%%%%%
%% MAINMATTER
%%%%%%%%%%%%%%%%%%%%%%%%%%%%%%%%%%%%%%%%%%%%

\section{INTRODUCTION AND BACKGROUND}

White dwarfs are the end state for over 95\% of all stars in the Galaxy, including our Sun, and 
their circumstellar environments represent the ultimate fate for essentially all planetary systems 
as well as the Solar System.  Only low-mass stars of spectral type K and M are more common
than white dwarfs both in the Solar neighborhood\footnote{\url{http://www.recons.org}} as well
as the Galaxy at large.  The disk white dwarfs seen today are primarily the descendants of A- 
and F-type stars with masses $1.2-3.0$\,$M_{\odot}$, although their population contains the 
remnants of all intermediate-mass stars that avoid core collapse.  

Hence white dwarf are {\em evolved but not necessarily old}.  The nearest white dwarf to the
Sun is Sirius\,B with a total age near 250\,Myr \citep{lie05}, while the Pleiades, Hyades, and
Praesepe young open clusters together contain more than one dozen white dwarf members.  
The low and typically blue-peaked luminosities of white dwarfs make them excellent targets 
for direct imaging planet searches and substellar companion studies in general.

\subsection{White Dwarfs as Heavy Element Detectors}

Surface gravities of $\log\,g\,[({\rm cm\,s}^{-2})]=8$, assisted by the onset of convection, 
ensure that heavy elements sink rapidly in the atmospheres of white dwarfs as they cool below 
25\,000\,K \citep{koe09,paq86}.  From this point forward in evolution, the timescales for metals 
to diffuse below the photosphere are always a few to several orders of magnitude shorter than the 
cooling age.  Thus, cool white dwarfs should have atmospheres composed of pure hydrogen or 
helium, an expectation corroborated by observation \citep{eis06}.

Interestingly, a significant fraction of these stars display photospheric absorption lines due 
to metals when viewed with high-powered optical spectroscopy \citep{koe05,zuc03}.  These 
atmospheric heavy elements are external pollutants that imply ongoing accretion rates above
10$^8$\,g\,s$^{-1}$ or metal masses on the order 10$^{22}$ \,g within the convection zone of 
the star \citep{far10a,koe06}.  While mass can be accreted via Roche lobe overflow or stellar 
wind of a binary companion, for {\em single} white dwarfs with metals the two possibilities are 
the interstellar medium or circumstellar material.  Regardless of the source of photospheric
contamination, {\em the heavy element abundances in cool white dwarfs indirectly measure 
the composition of the accreted matter}.

\subsection{Two Important Polluted Prototypes}

\subsubsection{van Maanen 2}

\begin{figure}
\includegraphics[height=0.50\textheight]{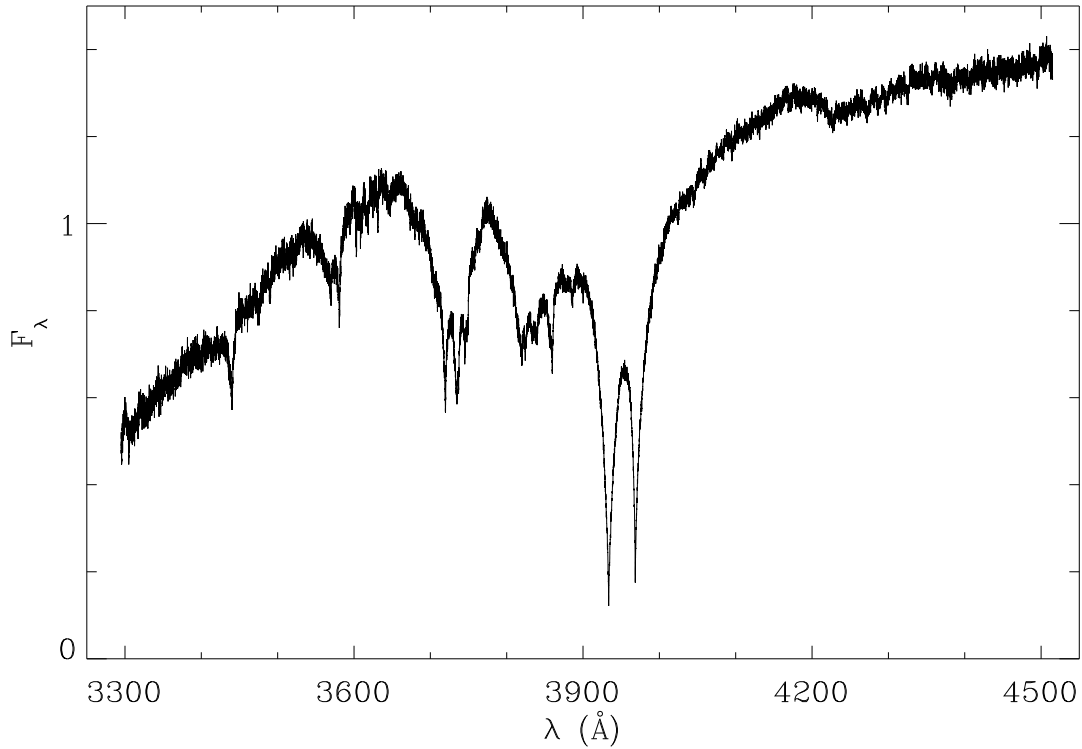}
\caption{vMa\,2 was the third white dwarf to be discovered \citep{van19}, and is the prototype of 
the DZ spectral class.  This $R\approx20\,000$ VLT UVES spectrum (which extends to 6700\AA) 
reveals only metallic lines of iron, calcium, and magnesium; the pollution is refractory-rich yet 
hydrogen deficient.  From the SPY project \citep{nap03}.}
\label{fig1}
\end{figure}

Figure \ref{fig1} shows the optical spectrum of the prototype metal-polluted white dwarf and 
nearest single degenerate star to the Sun, vMa\,2.  The prominent calcium and iron absorption 
features led \citep{van17} to initially conclude his high proper motion star was of early F-type.  
Only 40 years later was it understood that these remarkably strong features in a white dwarf 
were due to a metal abundance 30\,000 times lower than solar \citep{wei60}.  Thus, white 
dwarfs can be detectably polluted by a relatively small amount of metals.  The calcium K line is 
the trademark of metal-polluted white dwarfs, as this feature is detected in the optical spectra
of all members of this class.

vMa\,2 is a helium atmosphere white dwarf with metals.  Such stars have spectral type DBZ or DZ 
(D for degenerate star; B if warm enough to exhibit helium lines; Z for metallic features).  Owing to 
the relative transparency of helium, the first two dozen externally polluted white dwarfs discovered 
were of this same spectral class; hydrogen-poor stars enriched by metals \citep{sio90}.  Their 
hydrogen deficiency was the first and still most fundamental problem with the interstellar accretion
hypothesis.  Because hydrogen floats and heavy elements sink, the abundance pattern in DBZ 
stars is precisely the opposite of expectations if hydrogen-dominated interstellar matter were the
source of the atmospheric pollution.

\subsubsection{G29-38}

\begin{figure}
\includegraphics[height=0.45\textheight]{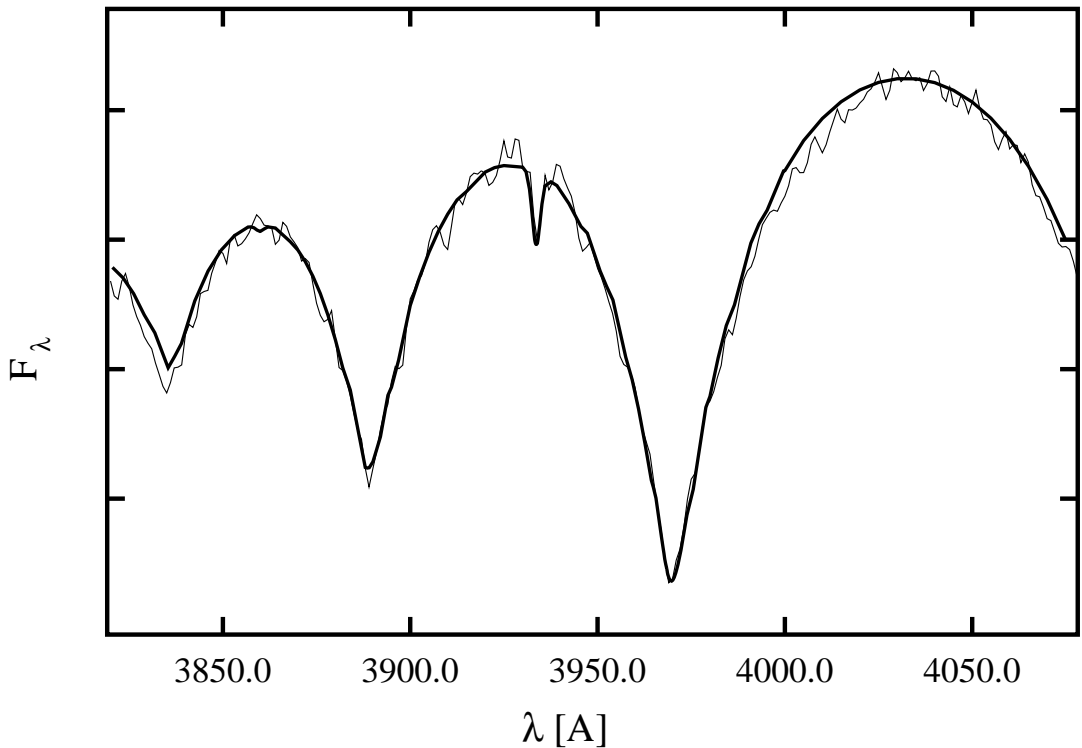}
\caption{The detection of the Ca II K-line in the optical spectrum of G29-38.  Note the much stronger 
line seen in vMa\,2 represents an abundance 1500 times lower than determined for G29-38.  Line 
strength is a strong function of atmospheric composition (i.e.\ hydrogen or helium) and stellar effective 
temperature.  From \citet{koe97}.}
\label{fig2}
\end{figure}

Figure \ref{fig2} displays the optical spectrum of the metal-rich white dwarf G29-38 \citep{koe97}.  
The relatively weak calcium K line in this star implies a metal abundance 1500 times higher than 
in vMa\,2, because G29-38 has a hydrogen atmosphere and corresponding spectral type DAZ (D 
for degenerate star; A for hydrogen lines; Z for metallic features).  Its ultraviolet spectrum exhibits 
numerous iron features and magnesium \citep{koe97}, the latter is also seen in a high-resolution
optical spectrum \citep{zuc03},

Although not the prototype of its spectral class, the importance of G29-38 rests upon the detection
of $T\approx1000$\,K circumstellar dust \citep{zuc87} ten years prior to the photospheric metals.  
Together, these observational data indicate the dust orbits sufficiently close to pollute the star via
infall, and corroborates dust temperature inferred from the observed infrared excess.  G29-38 is the 
prototype dusty white dwarf, and the only known instance of circumstellar dust and photospheric
metals prior to the launch of {\em Spitzer}.

\subsection{Metal-Contaminated White Dwarfs at Large}

Understanding externally polluted white dwarfs themselves is key to the nature of the source of
their contamination.  The helium-rich DBZ stars have been observed in significant numbers for
many decades, though mainly as a result of their atmospheric transparency.  Metals may persist in 
DBZ stars at detectable levels for 10$^6$\,yr timescales \citep{paq86}, and combined with their
$10^8-10^9$\,yr cooling ages, it is understandable the interstellar medium was initially the prime
culprit for their pollution \citep{dup93a,dup93b,dup92}.  In addition to the problem of their hydrogen
deficiency, there remained insufficient evidence for local interstellar clouds responsible for many 
nearby polluted stars \citep{aan93}.

The gradual discovery of numerous members of the DAZ spectral class \citep{zuc98,hol97,lac83}
provided new insights into the general issue of white dwarf pollution.  The heavy element sinking 
timescales in hydrogen atmosphere white dwarfs can be a short as {\em a few days} \citep{paq86}, 
implying the observed abundances in DAZ stars are maintained by {\em ongoing metal accretion}.
Inferred metal accretion rates for these stars are typically on the order of 10$^8$\,g\,s$^{-1}$ but
again the necessary, local interstellar clouds are distinctly lacking \citep{koe06,zuc03}.

The two major differences between the DAZ and DBZ stars is the size of the convection zone 
and the transparency of their atmospheres.  The former leads to the sizable difference in the 
metal diffusion timescales \citep{koe09} and the latter determines the detectability of metals for 
a given abundance.  To summarize:

\begin{itemize}

\item{DAZ white dwarfs have {\em hydrogen} atmospheres, {\em thin} convection zones, {\em 
short} diffusion timescales and hence accretion of heavy elements is inferred to be ongoing.}

\item{DBZ white dwarfs have {\em helium} atmospheres, {\em long} diffusion timescales, {\em 
deep} convection zones and thus contain large masses of heavy elements in their outer layers.}

\end{itemize}

\noindent
Combined with the difficulties with the interstellar accretion hypothesis, these two properties 
-- the large masses of metals already accreted by DBZ stars, and the ongoing metal accretion 
by the DAZ stars -- led to the generation of ground- and space-based programs to detect and
study circumstellar dust at metal-polluted white dwarfs.

\section{DUST DISK SEARCHES AND STUDIES}

While the detection of dust at G29-38 took place 17 years prior to the launch of {\em Spitzer}, 
it was during this latter era that a wealth of data and insight were achieved in the study of metal-$
$contaminated white dwarfs and their circumstellar environments.  The increased attention and 
research focussed on white dwarfs and circumstellar dust began soon after \citet{zuc03} published 
the first extensive and successful survey to detect nearby DAZ stars and \citet{jur03} published the 
tidally disrupted asteroid model for G29-38.

\subsection{Destroyed Minor Planets in the Post-Main Sequence}

Two influential models emerged between 2002 and 2003 that lent theoretical support to the idea
of circumstellar pollution in cool white dwarfs.  The first was a model of planetary system instabilities
introduced by post-main sequence mass loss.  \citet{deb02} demonstrated that following the bulk of 
mass loss during the asymptotic giant brach, the unperturbed semimajor axis ratios in multiple planet
systems are unchanged, but new resonances are established.  In essence, the planetary system
becomes dynamically young again for a timescale of 10$^8$\,yr, similar to proto-planetary systems 
and that inferred for the Solar System.

\citet{jur03} utilized this general picture of a dynamically renewed planetary system to account for
all the observed properties of G29-38 by modeling its dust disk as the result of a tidally destroyed
asteroid.  In this picture, a previously stable asteroid is perturbed into a high eccentricity orbit that
passes within the Roche limit of the white dwarf, where it is shredded by gravitational tides.  Via
collisions, the debris rapidly produces small dust particles orbiting in a vertically optically thick, 
geometrically thin (flat) disk.  The closely orbiting dust 1) emits in the infrared, producing the 
observed excess at G29-38, and 2) gradually falls onto the star, polluting its atmosphere with 
the observed heavy elements.  This is currently the standard model for metal-enriched white 
dwarfs with infrared excess.

\begin{table}
\begin{tabular}{ccccccc}

\hline

\tablehead{1}{c}{c}{WD}
& \tablehead{1}{c}{c}{Name}
& \tablehead{1}{c}{c}{Type}
& \tablehead{1}{c}{c}{{\em T$_{\rm \bf eff}$} (K)}
& \tablehead{1}{c}{c}{Year}
& \tablehead{1}{c}{c}{Telescope}
& \tablehead{1}{c}{c}{Reference\tablenote{(1) \citealt{zuc87};
(2) \citealt{bec05};
(3) \citealt{kil05};
(4) \citealt{kil06};
(6) \citealt{kil07};
(6) \citealt{von07};
(7) \citealt{jur07a};
(8) \citealt{far08b};
(9) \citealt{far09b};
(10) \citealt{bri09};
(11) \citealt{far10c};
(12) \citealt{mel10}
}} \\

\hline

2326$+$049      	&G29-38			&DAZ		&11700	&1987	&IRTF			&(1)\\
1729$+$371		&GD\,362			&DBZ		&10500	&2005	&IRTF/Gemini		&(2,3)\\
0408$-$041		&GD\,56			&DAZ		&14400	&2006	&IRTF			&(4)\\
1150$-$153		&EC\,11507$-$1519	&DAZ		&12800	&2007	&IRTF			&(5)\\
2115$-$560      	&LTT\,8452		&DAZ		&9700 	&2007	&{\em Spitzer}		&(6)\\
0300$-$013		&GD\,40			&DBZ		&15200	&2007	&{\em Spitzer}		&(7)\\
1015$+$161		&PG				&DAZ		&19300	&2007	&{\em Spitzer}		&(7)\\
1116$+$026		&GD\,133			&DAZ		&12200	&2007	&{\em Spitzer}		&(7)\\
1455$+$298		&G166-58			&DAZ		&7400	&2008	&{\em Spitzer}		&(8)\\
0146$+$187		&GD\,16			&DBZ		&11500	&2009	&{\em Spitzer}		&(9)\\
1457$-$086      	&PG				&DAZ		&20400	&2009	&{\em Spitzer}		&(9)\\
1226$+$109		&SDSS\,1228 		&DAZ		&22200	&2009	&{\em Spitzer}		&(10)\\
0106$-$328		&HE\,0106$-$3253	&DAZ		&15700	&2010	&{\em Spitzer}		&(11)\\
0307$+$077		&HS\,0307$+$0746 	&DAZ		&10200	&2010	&{\em Spitzer}		&(11)\\
0842$+$231		&Ton\,345 		&DBZ		&18600	&2008	&{\em AKARI} 		&(11)\\
1225$-$079:		&PG				&DBZ		&10500	&2010	&{\em Spitzer}		&(11)\\
2221$-$165		&HE\,2221$-$1630 	&DAZ		&10100	&2010	&{\em Spitzer}		&(11)\\
1041$+$091		&SDSS\,1043 		&DAZ		&18300	&2010	&CFHT/Gemini		&(12)\\

\hline
\end{tabular}
\caption{The First 18 White Dwarfs with Infrared Excess Due to Circumstellar Dust}
\label{tbl1}
\end{table}

\subsection{First {\em Spitzer} Observations of White Dwarfs with Dust and Metals}

\begin{figure}[h!]
\includegraphics[height=0.86\textheight]{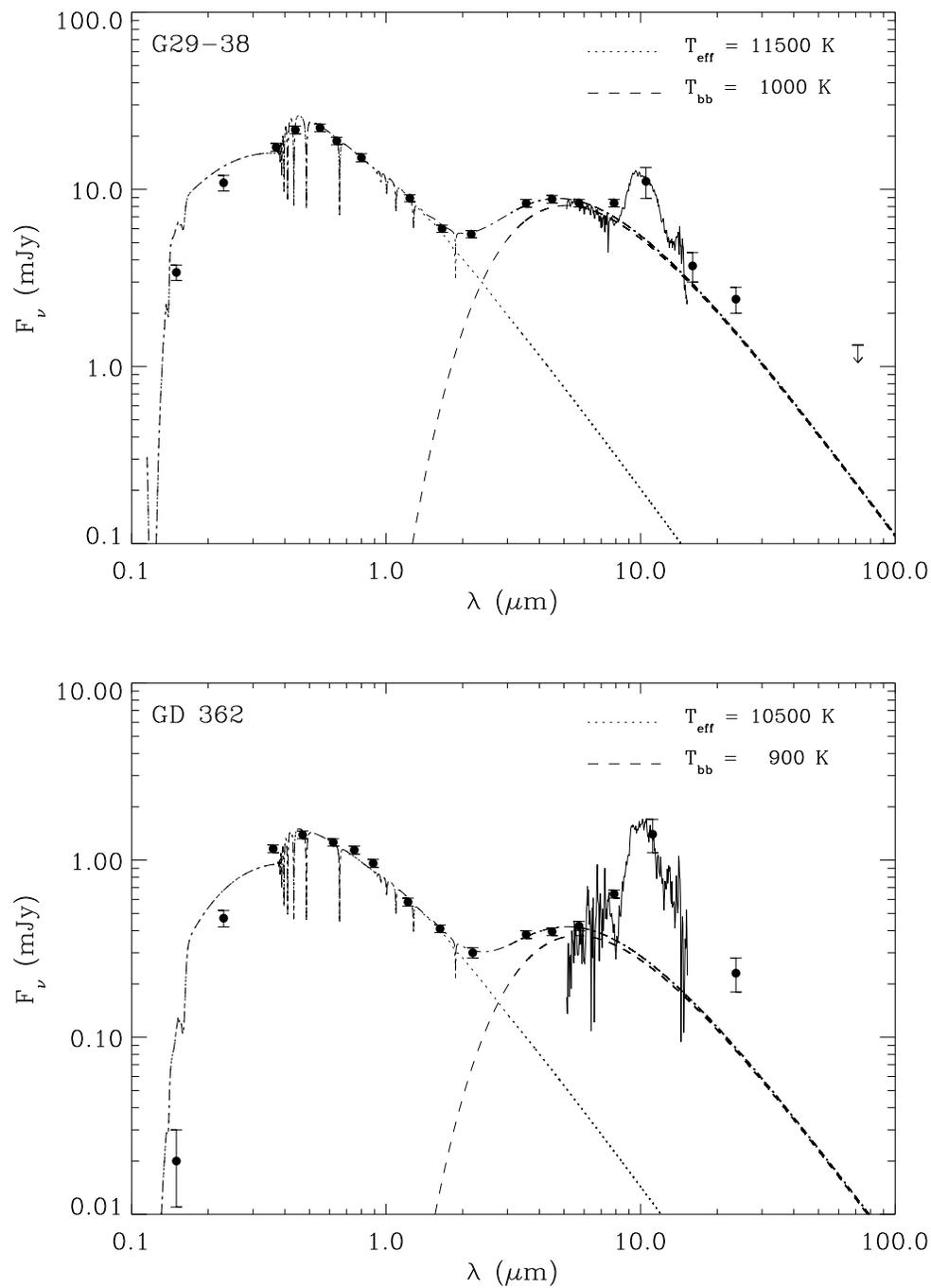}
\caption{The broad SEDs of G29-38 and GD\,362.  The data points with error bars show short
wavelength and {\em Spitzer} photometry fitted by 1) stellar atmosphere models drawn as dotted
lines, and 2) thermal blackbody models shown as dashed lines.  Also shown are their {\em Spitzer} 
infrared spectra revealing strong silicate emission from olivine particles \citep{jur07b,rea05}.}
\label{fig3}
\end{figure}

The first two white dwarfs found to have circumstellar dust disks were naturally premier targets for 
{\em Spitzer}.  Figure \ref{fig3} shows the full spectral energy distributions (SEDs), including infrared 
data obtained by {\em Spitzer} for G29-38 and the second white dwarf found to have circumstellar dust, 
GD\,362 \citep{bec05,kil05}.  It is noteworthy that the strength and shape of their thermal continua
are similar, with a very warm temperature excess and no cool component.

Both stars have strong emission features due to silicate dust particles, specifically olivines similar
to in their emission to solids found in the zodiacal cloud and orbiting stars like $\beta$\,Pic where 
planet formation is ongoing \citep{jur07b,rea05}.  Compared to G29-38, the infrared emission from 
GD\,362 is spectacular and perhaps the strongest silicate emission ever seen at a mature star
\citep{son05}.  The remarkable infrared properties of GD\,362 are matched by its highly polluted 
atmosphere (\citealt{zuc07}; discussed below).

Importantly, the infrared emission of both stars is well-modeled by the optically thick, flat disk 
models of \citet{jur03}, placing all the circumstellar material within 1\,$R_{\odot}$ of the white
dwarf.  At this distance, any rocky parent body larger than a few km in size would become tidally
disrupted \citep{dav99}, making the observations consistent with a destroyed minor planet.

\subsection{Disk Properties: Infrared Photometry and Spectroscopy}

As of this writing (mid-2010), there are 18 metal-contaminated white dwarfs known to have an infrared 
excess owing to circumstellar dust as confirmed via {\em Spitzer} mid-infrared observations \citep{far10c,
far09b,bri09,far08b,von07,jur07a,jur07b,rea05}.  Table \ref{tbl1} lists these white dwarfs with dust disks 
in order of the discovery by publication date, together with the stellar effective temperature, spectral type,
and the telescope that first detected the excess emission.  As can be seen from the Table, {\em Spitzer}
has been instrumental in this field.

More than 60 metal-rich white dwarfs have been observed with {\em Spitzer} IRAC at wavelengths 
between 3 and 8\,$\mu$m, and roughly half were observed with MIPS at 24\,$\mu$m before the end of
the cryogenic lifetime \citep{far09b}.  Eight white dwarfs with disks were observed spectroscopically with
{\em Spitzer} IRS between 5 and 15\,$\mu$m (representing all observable white dwarfs with disks at the 
time).  From this large dataset, it is found that all dust disks detected at white dwarfs share a common set
of characteristics, supporting the idea that a single mechanism is responsible for their appearance.

\begin{enumerate}

\item{Circumstellar dust at white dwarfs is well-modeled by geometrically thin, flat disks that are
vertically optically thick at wavelengths as long as 20\,$\mu$m.}

\item{Infrared emission from dust at white dwarfs is very warm, with the inner disk at temperatures
($T\sim1200$\,K) where grains rapidly sublimate.}

\item{The outer disk radius is always constrained to lie within the tidal breakup limit for km-size or 
larger bodies; there is no evidence for cool dust at white dwarfs.}

\item{{\em Minimum} dust masses are 10$^{18}$\,g, but because the disks are optically thick (and 
based on additional evidence) the disks may have masses as large as 10$^{24}$\,g.}

\item{Infrared spectroscopy reveals the orbiting material is rich in silicate minerals (olivine primarily
but also possibly pyroxene), and deficient in carbon and hydrogen.}

\item{The stellar pollution caused by the infall of disk material is rich in refractory and transitional
metals, but poor in volatile elements.}

\end{enumerate}

The observed properties provide a few important constraints on the origin and behavior of the 
circumstellar material.  First, the lack of cool dust at white dwarfs implies the material has not been 
captured from the interstellar medium.  Such material would migrate inward from the Bondi-Hoyle 
radius (roughly 1\,AU for a typical white dwarf), but the compact nature of these disks ($r<0.01$\,AU) 
rules out material at these distances \citep{far09b,jur07a}.  Second, the material is depleted in volatile
elements, such as carbon, relative to metals and similar to matter found within the inner Solar System 
\citep{jur09a,far08b,jur06,rea05,lod03}.  Third, the dust disks persist all the way up to temperatures
where dust should rapidly sublimate, precisely as expected for material that is delivering heavy 
elements to the photosphere of its white dwarf.

In summary, {\em the debris orbiting white dwarfs is rocky.  The gradual pollution of the stellar
atmosphere by the circumstellar material provides a powerful tool to measure the bulk chemical
composition of extrasolar, rocky planetary bodies such as asteroids, moons, and potentially major
planets}.

\begin{figure}
\includegraphics[height=0.45\textheight]{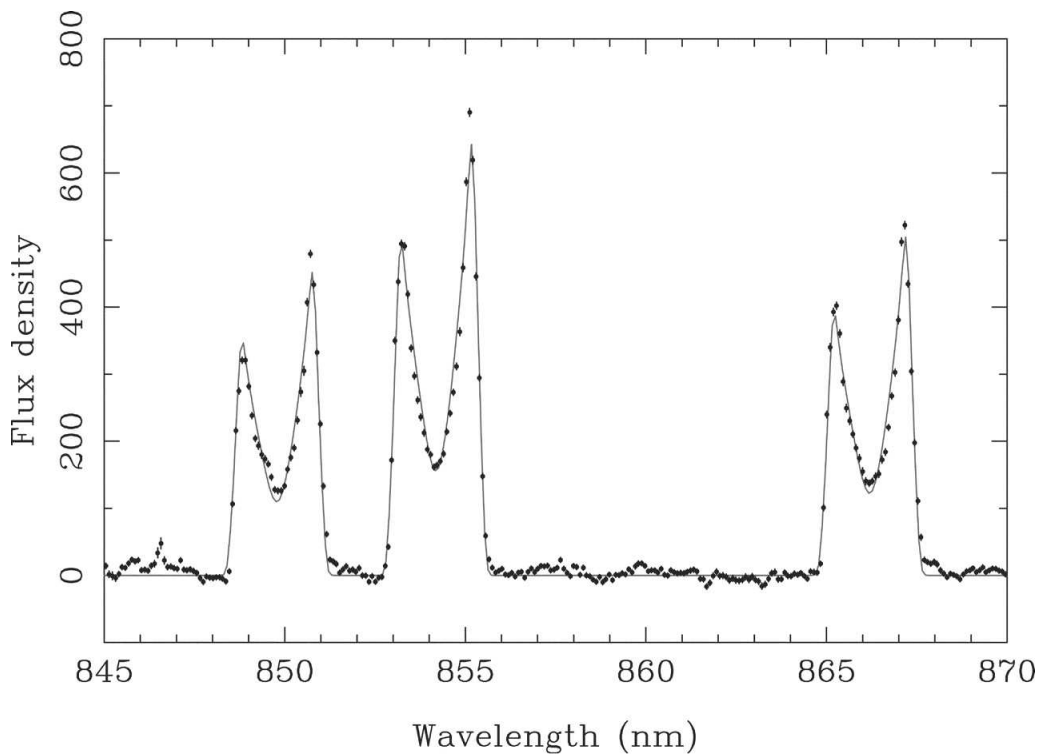}
\caption{Calcium emission in the optical spectrum of SDSS\,1228.  The width of the emission
lines indicates Keplerian rotation at $\pm$ 630\,km\,s$^{-1}$.  Emission lines from iron are also
seen in this star, but not from hydrogen or helium.  From \citet{gan06}.}
\label{fig4}
\end{figure}

\subsection{Disks with Metallic Gas Components}

There are three white dwarfs with dust and photospheric metals that were initially discovered due
to the presence of metallic, {\em gaseous} emission lines in the optical \citep{gan08,gan07,gan06}.
Found serendipitously in the Sloan Digital Sky Survey (SDSS), each of these white dwarfs shows 
optically thick emission lines in the calcium triplet in the far-red region of their optical spectra.  
Furthermore, these emission lines are precisely the type that are seen in the gas-dominated accretion 
disks of cataclysmic variables, but in these {\em single} white dwarfs with disks, hydrogen and helium 
are seen only in absorption \citep{gan08}.  Therefore, the emitting material is solely composed of heavy
elements.

Importantly, the gaseous, emitting disk components yield empirical constraints on their radial location,
placing them firmly within about 1.2\,$R_{\odot}$.  This comes directly from Kepler's laws applied to
the observed velocity broadening of the emission features \citep{gan06}.  While the modeling of the
dust emission in the infrared indicates similar radial distributions for disks at white dwarfs, these stars
with metallic emission corroborate and strengthen the idea that disks at cool white dwarfs result from
the tidal destruction of asteroids.

The gaseous debris in these systems coexists with dust, and is part of the same phenomenon as 
the dusty white dwarfs where calcium emission lines are not seen \citep{mel10}.  That is, the gas is
not the result of sublimated dust very close to the white dwarfs, as the dust and gas occupy the same 
circumstellar regions \citep{bri09}.  Rather, the gas is the result of energetic collisions within a rapidly 
rotating disk with speeds exceeding $0.001c$ -- if the velocity dispersion among dust particles is only
1\%, then collisions at 6\,km\,s$^{-1}$ will occur and grind solids into gas \citep{jur08}.

\subsection{Disk Properties: Stellar Spectroscopy}

Because cool white dwarfs should have pure hydrogen or helium atmospheres, spectroscopy of 
the metal-contaminated stars indirectly yields the elemental abundances in the accreted material.
\citet{jur06} showed that several metal-lined white dwarfs have iron-to-carbon ratios that are super 
solar, indicating the material originated in a high temperature region of its host system (i.e.\ the
inner system).  Furthermore, the two published cases where detailed abundance analyses were 
obtained high-powered spectroscopy reveal material broadly similar to the terrestrial planets 
\citep{kle10,zuc07}.  

\begin{figure}
\includegraphics[height=0.45\textheight]{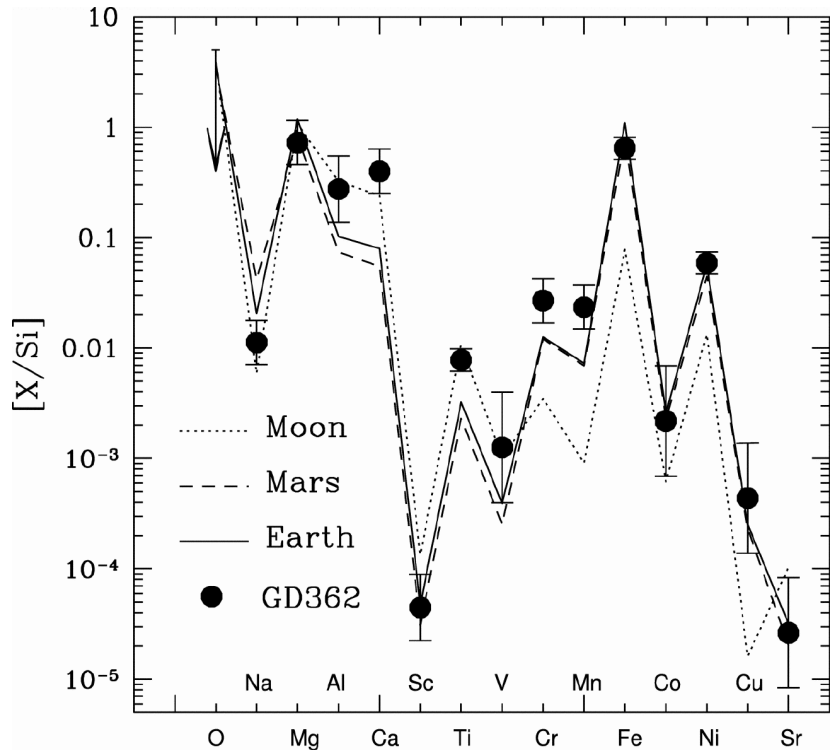}
\caption{Fifteen elements heavier than helium in the highly contaminated photosphere of GD\,362
reveal a pattern that is similar to a combination of Earth and Moon compositions.  From \citet{zuc07}.}
\label{fig5}
\end{figure}

Not only are the compositions of metal-enriched white dwarfs distinctly rocky and indicative of
asteroidal or terrestrial debris, but so are the metal masses.  In helium-rich DBZ stars such as GD\,362 
and GD\,40, the total mass of heavy elements in the mixing layer of the stars is greater than 10$^{22}
$\,g; more material than contained in a typical 200\,km Solar System asteroid(!).  However, this is only 
the mass of metals currently within the outer layers, and hence the {\em minimum} mass of the accreted 
matter.  If the star has possessed a disk for a few to several diffusion timescales (rather likely in the case 
of GD\,40 \citealt{kle10}), then the total mass of heavy elements involved is approaches the masses of 
Vesta and Ceres, the two largest asteroids in the Solar System.  In the case of GD\,362, in order to 
account for all of its spectacular properties with a single pollution event, \citet{jur09b} find that the 
necessary parent body mass is between that of Jupiter's moon Callisto and Mars, including {\em 
evidence for water}.

\section{A Picture of Planetary System Remnants}

The {\em Spitzer} and ground-based detections of circumstellar dust are useful in that they support 
an emerging picture of rocky planetary system remnants, and provide a physical and chemical link 
between the orbiting debris and the photospheric heavy elements.  In this manner, astronomers now
have a powerful tool to study the composition of extrasolar terrestrial bodies.  But the phenomenon
of destroyed and accreted minor planets at white dwarfs needs an evolutionary context, and there
are still significant uncertainties in how specific properties arise.

\subsection{Disk Properties: Frequency and Statistics}

\begin{figure}[h!]
\includegraphics[height=0.50\textheight]{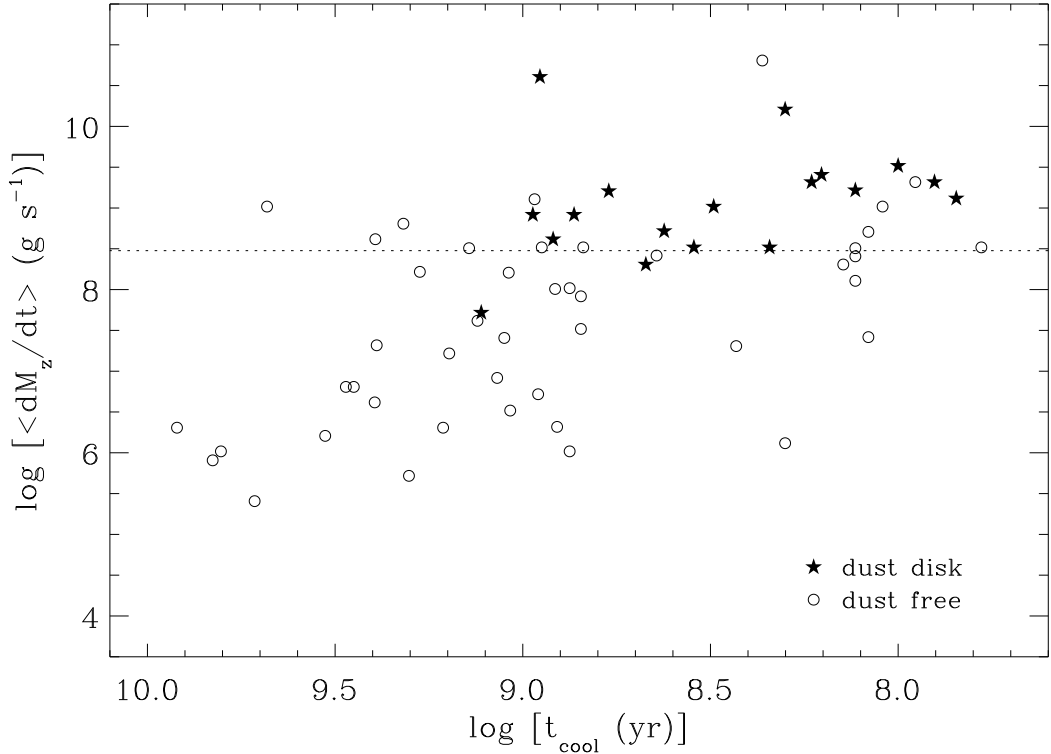}
\caption{Dust disk frequency among all 61 metal-rich white dwarfs observed by {\em Spitzer}
IRAC.  Plotted on are the time-averaged metal accretion rate and cooling age for each star. The 
dotted line corresponds to $3\times10^8$\,g\,s$^{-1}$.  G166-58 is the only star with a disk that 
is located significantly below this accretion rate benchmark, and with a cooling age beyond 1\,Gyr
\citep{far08b}.  Adapted from \citet{far10c}.}
\label{fig6}
\end{figure}

After three cycles of {\em Spitzer} studies of DAZ and DBZ white dwarfs, there were sufficient stars
observed to begin robust statistics.  \citet{far09b} assembled and analyzed the first 53 metal-rich white
dwarfs, finding that the likelihood of a circumstellar dust was strongly linked with the metal accretion
rate of the star and its cooling age.  The time-averaged metal accretion rate is calculated by multiplying
the metal abundance in the star by its convective envelope mass, and dividing by the metal sinking
timescale.  This essentially assumes a steady-state balance between accretion of heavy elements
from an external source and their diffusion below the outer layers \citep{koe09,koe06}.  For the DAZ
stars whose diffusion timescales are very short, the assumption of ongoing accretion is physically
motivated, while for the DBZ stars the time-averaged accretion rate is simply a valuable diagnostic
\citep{far09b}.

Figure \ref{fig6} plots the inferred metal accretion rates versus cooling age for all polluted white 
dwarfs observed by {\em Spitzer} to date, and distinguishes between those stars with infrared excess 
and those without.   As can be seen in the plot, for accretion rates above $3\times10^8$\,g\,s$^{-1}$, 
there is a greater than 50\% chance a white dwarf will have an infrared excess due to dust.  This makes 
physical sense because the greater the ongoing accretion rate, the (presumably) more massive the 
reservoir that is supplying the metals.  This trend is also consistent with the idea that more massive 
disks should have a sufficiently high surface density to eschew being mostly or totally vaporized via 
collisions \citep{jur08}.  In essence, massive disks are tightly packed, with the inter-particle spacing 
of the same order as the particle size, effectively damping out collisions \citep{far08b}.  

Another observed trend is that dust disk frequency correlates with younger cooling ages.  However, 
this trend contains a bias because metals can only be detected warmer and younger white dwarfs 
when the pollution is relatively high.  Still, there is only a single white dwarf with atmospheric metals 
and an infrared excess at ages beyond 1\,Gyr \citep{far10c}.

When one views these metal-rich stars statistically, as being drawn from larger samples of white 
dwarfs in general, the overall result is that between 1\% and 3\% of all white dwarf with cooling 
ages less than around 0.5\,Gyr have both metal-enriched atmospheres and circumstellar dust.  
Such a picture is consistent with the dynamical resettling of a planetary system in the post-main 
sequence \citep{deb02}, and the relative dearth of disks at older cooling ages may represent the
depletion of large asteroids necessary to create an infrared excess \citep{far09b}.  Minimally at
the few percent level, white dwarfs with dust represent an underlying population of stellar systems 
harboring the remnants of terrestrial planets \citep{far09b}, and therefore at least this fraction of 
A- and F-type stars are likely to build rocky planets.

\subsection{Narrow Dust Rings or Largely Gaseous Disks}

Figure \ref{fig6} also demonstrates that the majority of observed metal-rich white dwarfs do not
exhibit infrared excess detectable by {\em Spitzer}, including some of the most highly polluted stars.
Prior to the launch of {\em Spitzer} it was suggested that unseen, low-mass stellar and brown dwarf 
companions could be responsible for the metal pollution seen in some white dwarfs, e.g.\ via wind 
capture \citep{dob05,hol97}.  However, the mid-infrared photometry and spectra of metal-lined white
dwarfs rules out the presence of low-mass companions down to 25\,$M_{\rm Jup}$ and in some 
cases to even lower substellar masses \citep{far08a,bar03}.  Where found, the detected infrared 
excesses cannot be reproduced by substellar atmosphere models, while the majority of metal-rich
stars simply have no infrared excess where one would be expected if a companion were present
\citep{far09b}.

A realistic possibility for those stars with relatively long metal sinking timescales (DBZ and very 
cool DAZ stars) is a depleted disk.  In this case, the extant photospheric metals are the scars of a 
previous accretion event that took place within a few diffusion timescales \citep{far09b}.  In these
cases one might expect to see abundances that reflect this fact, as iron sinks most rapidly, while
lighter metals such as sodium and magnesium diffuse more slowly \citep{koe09}.

However, there are many cases of white dwarfs where metal accretion must be ongoing due to
their very short diffusion timescales \citep{koe06}.  For these stars where an infrared excess is 
not detected, accretion is occurring but not being detected.  If collisions within an evolving disk 
typically grind down solids into particles too small ($2\pi a / \lambda \ll 1$) to emit efficiently at 
infrared wavelengths, then a largely or totally gaseous disk may result \citep{far08b}.  Furthermore, 
asteroid destruction within the Roche limit of a white dwarf should occur more frequently for small
asteroids as they should be more populous and their low mass implies they will be more readily 
perturbed into an eccentric orbit.  Such multiple impacts on a pre-existing disk will strongly enhance 
collisions and sputtering and should result in substantial gaseous circumstellar material \citep{jur08}.

\begin{figure}[h!]
\includegraphics[height=0.86\textheight]{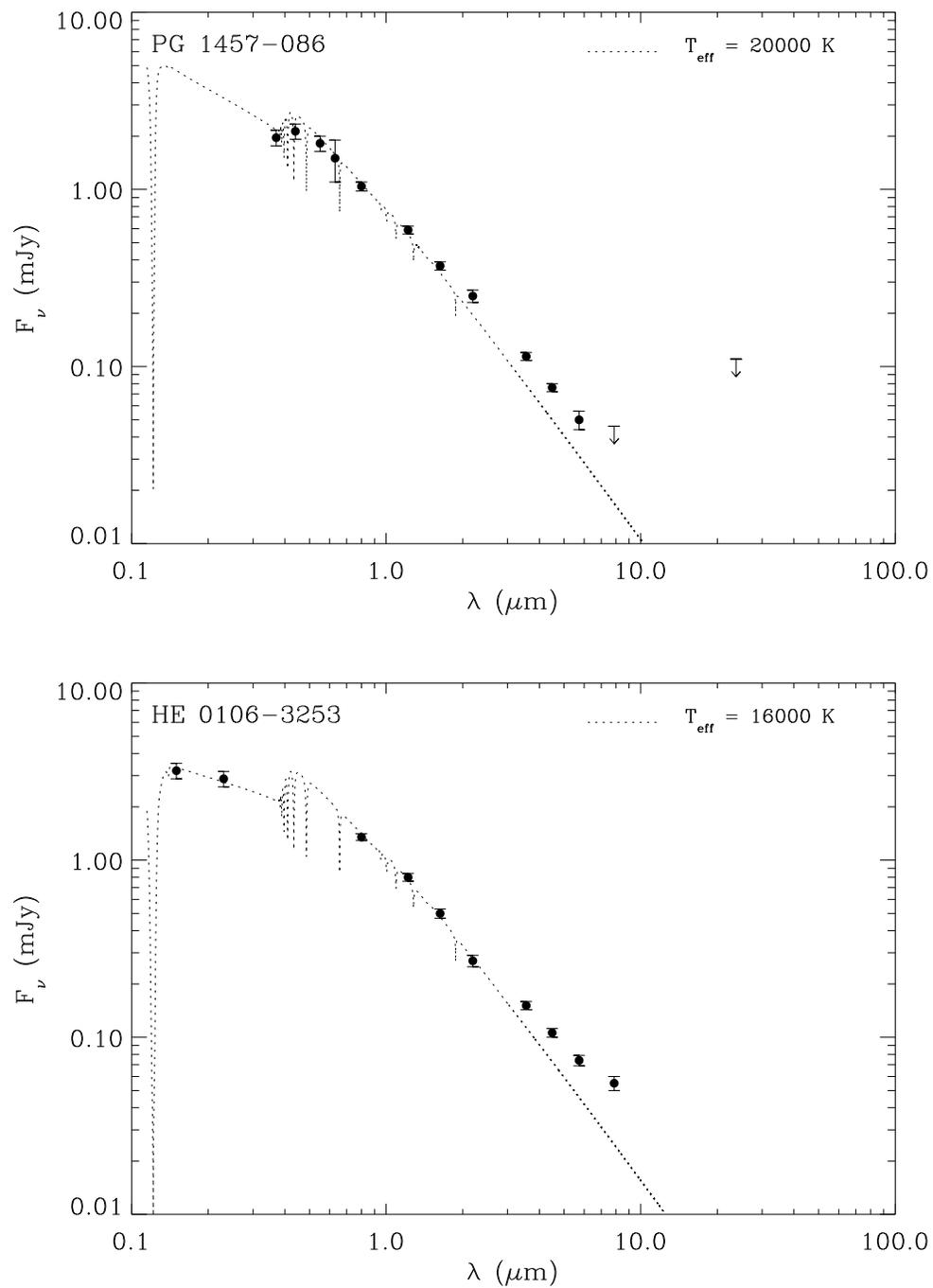}
\caption{The two most subtle infrared excesses detected from narrow, circumstellar dust rings
\citep{far10c,far09b}.  Without supporting short wavelength data, the infrared photometry would have
been difficult to interpret as an excess with confidence.}
\label{fig7}
\end{figure}

Another possibility is that many white dwarfs have circumstellar dust from which they are accreting, 
but the infrared signature of the disk is subtle or undetectable.  Figure \ref{fig7} shows the relatively
subtle infrared excesses of two metal-rich white dwarfs observed by {\em Spitzer}.  In each case, the
excess is in the $3-5\sigma$ range, whereas a typical dust disk like G29-38 is detected at $15-20
\sigma$ \citep{far10c}.  Importantly for the physics involved, narrow rings with radial extents in the
range $\Delta r = 0.01 - 0.1$\,$R_{\odot}$ are possible but would not necessarily produce an excess
unless viewed at a modest inclination angle.  Yet dust rings this narrow would still contain sufficient
mass to pollute a typical white dwarf for $10^5 - 10^6$\,yr at the accretion rates inferred for highly
polluted stars \citep{far10c}.

In summary, those stars with infrared excess and dust may be the result of a single, large asteroid
and hence spectroscopy of their polluted atmospheres will yield the {\em composition of a planetary 
embryo or fragment}.  For those stars without obvious infrared excess, multiple events may have 
caused the debris to become largely gaseous, and spectroscopy of these stars may reveal the 
{\em composition of an ensemble of smaller asteroids}.

\begin{figure}
\includegraphics[height=0.50\textheight]{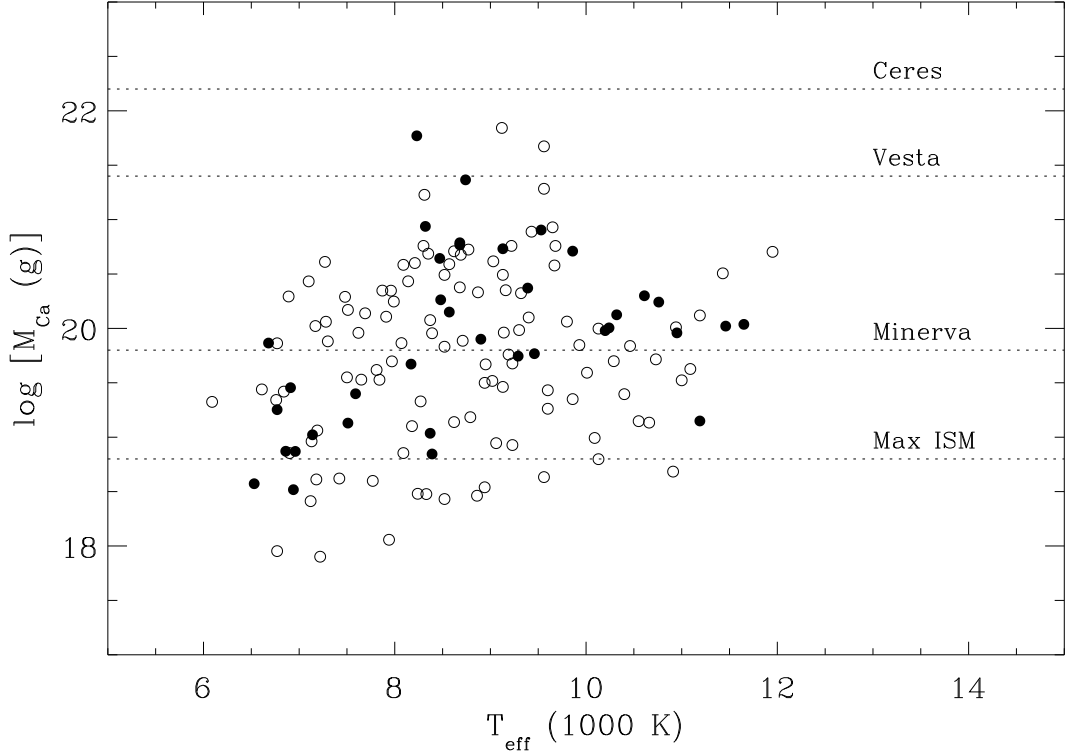}
\caption{Calcium masses in the convective envelopes of 146 DBZ stars from the SDSS. The open 
and filled circles represent stars with and without trace hydrogen, respectively.  The top three dotted 
lines mark the mass of calcium contained in the two largest Solar System asteroids Ceres and Vesta, 
and the 150\,km diameter asteroid Minerva, assuming calcium is 1.6\% by mass as in the bulk Earth 
\citep{all95}. The dotted line at the bottom is the maximum mass of calcium that can be accreted over 
10$^6$\,yr by a cool white dwarf moving at a velocity of 50\,km\,s$^{-1}$ through an interstellar cloud 
with a density of 1000\,cm$^{-3}$.  Adapted from \citet{far10a}.}
\label{fig8}
\end{figure}

\subsection{The End of the Interstellar Accretion Hypothesis}

The accumulated evidence of the past several years strongly supports the idea that white dwarfs
with metals are polluted by their immediate circumstellar environments, and not by the interstellar
medium.  In fact, there is a distinct lack of evidence favoring the interstellar medium.

A recent analysis of nearly 150 metal-polluted (DBZ-type) white dwarfs discovered in the SDSS
used several diagnostics to test the feasibility of interstellar accretion for stars generally far more 
distant than previous studies.  No correlation was found between the metal content of the stars 
and their space velocities, as expected for Bondi-Hoyle type accretion, nor with their distance from
the Galactic plane, as expected if the stars are capturing material from within the spiral arms of the
Galaxy \citep{far10a}.  Furthermore, many of these stars are located far above the $\pm$ 100\,pc
thick gas and dust layer of the Galaxy, having spent several to tens of Myr outside the dominant
source of interstellar matter.  Figure \ref{fig8} perhaps illustrates best that interstellar accretion
simply cannot account for the metal content of these stars.  Simply stated, the white dwarfs contain
large asteroid-size masses of heavy elements that cannot have been accreted from the ISM by any 
reasonable physical model \citep{far10c}.

Thus, interstellar accretion is no longer viable.  Specifically, it cannot account for the observed 
population of metal-enriched white dwarfs.  Circumstellar pollution -- planetary system remnants --
is currently the only plausible and substantiated model.

\section{Outlook in a New Paradigm}

\subsection{Evidence for Water-Rich Asteroids}

The now complete transition to the planetary systems as the standard model for metals in white 
dwarfs suggests that other models for the origin and evolution of cool white dwarf atmospheres 
may benefit from reexamination.  For example, the hydrogen deficiency of helium atmosphere white 
dwarfs (both with and without metals) has always been a problem, as accretion from the interstellar 
medium should quickly convert them to hydrogen-rich atmospheres \citep{koe76}.  In the planetary 
system view, trace hydrogen can be delivered to the white dwarf in water-rich asteroids.  

The main asteroid belt in the Solar System is at least 6\% water by mass (due to the 25\% water 
content of Ceres; \citealt{tho05}), and it is reasonable to suppose that extrasolar asteroids will contain 
water.  While surface and near-surface water ice and volatiles will be evaporated due to heating during 
the asymptotic giant evolution of the white dwarf progenitor, interior water should survive in moderate to 
large size asteroids \citep{jur10}.  If one examines the relative fractions of cool, helium-rich white dwarfs
with 1) trace metals, 2) trace hydrogen, or 3) both, then the last category should occur with the lowest
frequency if the origin of the trace metals and the trace hydrogen are independent.  In contrast, trace to
these expectations, trace hydrogen is found more often in those stars with trace metals, indicating the
possibility of water within the parent body that delivered the metals \citep{far10a}.

There are a few examples of white dwarfs where the accretion of water-rich planetary material seems
likely.  Both GD\,362 and GD\,16 have circumstellar disks, photospheric metals, and rather high trace 
hydrogen abundances compared to other stars of the same effective temperature \citep{jur09b}.  A
dedicated search for atmospheric oxygen at these two stars is sure to be revealing.  The white dwarfs
GD\,378 and GD\,61 both have photospheric metals and oxygen detected in the ultraviolet, where the
oxygen abundance is in excess of that expected if all the metals were contained in oxides \citep{jur10}.
Because oxygen sinks more slowly than elements such as silicon, magnesium, and iron \citep{koe09},
the apparent overabundance of oxygen may be superficial.  If accretion can be shown to be ongoing 
in either of these systems, it would be strong evidence that these white dwarfs accreted a rocky parent
body rich in water.

\subsection{Terrestrial Planets at Intermediate Mass Stars}

White dwarfs are the evolutionary end point for the vast majority of all stars in the Milky Way.  Owing 
to the shape of the initial mass function and the finite age of the Galaxy, the currently observed disk 
population is primarily descended from main-sequence A- and F-type stars with masses in the range 
$1.2-3.0$\,$M_{\odot}$.  The statistics for tidally disrupted asteroids at white dwarfs can be used as a 
strict lower limit for the frequency of terrestrial planets at their main-sequence progenitors.  In this view, 
A- and F-type stars build terrestrial planets {\em at least} a few percent of the time, and in some cases 
there is evidence for water-rich building blocks.  If one attributes all metal-polluted white dwarfs to rocky 
debris, then the fraction of terrestrial planetary systems that survive post-main sequence evolution (at 
least in part) is as high as 20\% to 30\% \citep{zuc10,zuc03}.  Hence, the rocky planetary system 
remnants being witnessed at white dwarfs indicates that terrestrial planets are a frequent by-product 
of intermediate-mass star formation.

\bibliographystyle{aipproc}   % if natbib is available


\end{document}